\documentstyle[preprint,revtex,eqsecnum]{aps}
\begin{document}
\draft
%%%%%%%%%%%%%%%%%%%%%%%%%%%%%%%%%%%%%%%%%%%%%%%%%%%%%%%%%%%%%%%%%%%%%%
\begin{title}
Dirty blackholes:\\
Thermodynamics and horizon structure
\end{title}
\author{Matt Visser\cite{e-mail}}
\begin{instit}
Physics Department, Washington University, St. Louis,
Missouri 63130-4899
\end{instit}

\receipt{20 March 1992}
\begin{abstract}

Considerable interest has recently been expressed in (static
spherically symmetric) blackholes in interaction with various
classical matter fields (such as electromagnetic fields, dilaton
fields, axion fields, Abelian Higgs fields, non--Abelian gauge
fields, {\sl etc}). A common feature of these investigations that
has not previously been remarked upon is that the Hawking temperature
of such systems appears to be suppressed relative to that of a
vacuum blackhole of equal horizon area. That is: $k T_H \leq
\hbar/(4\pi r_H) \equiv \hbar/\sqrt{4\pi A_H}$.  This paper will
argue that this suppression is generic. Specifically, it will be
shown that
\[ k T_H = {\hbar\over4\pi r_H} \; e^{-\phi(r_H)} \;
          \left( 1 - 8\pi G \; \rho_H \; r_H^2 \right).  \]
Here $\phi(r_H)$ is an integral quantity, depending on the distribution
of matter, that is guaranteed to be positive if the Weak Energy
Condition is satisfied. Several examples of this behaviour will be
discussed.  Generalizations of this behaviour to non--symmetric
non--static blackholes are conjectured.

\end{abstract}

%%%%%%%%%%%%%%%%%%%%%%%%%%%%%%%%%%%%%%%%%%%%%%%%%%%%%%%%%%%%%%%%%%%%%%%%%%
\narrowtext
%%%%%%%%%%%%%%%%%%%%%%%%%%%%%%%%%%%%%%%%%%%%%%%%%%%%%%%%%%%%%%%%%%%%%%%%%%
\newpage
%%%%%%%%%%%%%%%%%%%%%%%%%%%%%%%%%%%%%%%%%%%%%%%%%%%%%%%%%%%%%%%%%%%%%%%%%%
\section{INTRODUCTION}
%%%%%%%%%%%%%%%%%%%%%%%%%%%%%%%%%%%%%%%%%%%%%%%%%%%%%%%%%%%%%%%%%%%%%%%%%%

For a variety of reasons, considerable attention has recently been
focussed on static spherically symmetric blackholes in interaction
with various static spherically symmetric classical fields. For
example, the system (gravity + electromagnetism + dilaton) has been
discussed by Gibbons and Maeda \cite{Gibbons-Maeda}, by Ichinose
and Yamazaki \cite{Ichinose-Yamazaki1989,Ichinose-Yamazaki1992},
and in an elegant paper by Garfinkle, Horowitz and Strominger
\cite{GHS}, this particular system currently being deemed to be of
interest due to its tentative connection with low energy string
theory. The resulting charged dilatonic blackholes were rapidly
generalized by Shapere, Trivedi, and Wilczek \cite{STW} to the
dyonic dilatonic blackholes appropriate to the system (gravity +
electromagnetism + dilaton + axion). The system (gravity +
electromagnetism + axion) has been considered by Allen, Bowick,
and Lahiri \cite{ABL}, by Campbell, Kaloper, and Olive \cite{CKO},
and by Lee and Weinberg \cite{Lee-Weinberg}.  The considerably
simpler system of (gravity + axion) and the associated axionic
blackholes had previously been discussed by Bowick, Giddings,
Harvey, Horowitz, and Strominger \cite{BGHHS}.  The system (gravity
+ electromagnetism + Abelian Higgs field) has been discussed by
Dowker, Gregory, and Traschen \cite{DGT} using Euclidean signature
formalism. Coloured blackholes, arising in the system (gravity +
non--Abelian gauge field), have been discussed by Galtsov and Ershov
\cite{Galtsov-Ershov}, by Straumann and Zhou \cite{Straumann-Zhou},
by Bizon \cite{Bizon}, and by Bizon and Wald \cite{Bizon-Wald}. A
variation on these themes: the system (gravity + axion + non--Abelian
gauge field), has recently been considered by Lahiri \cite{Lahiri}.
For brevity, any blackhole in interaction with nonzero classical
matter fields will be refereed to as ``dirty''.

A common feature of these various investigations is that whenever
the Hawking temperature of the resulting dirty blackhole can be
computed, the Hawking temperature (equivalently, the surface gravity)
appears to be suppressed relative to that of a clean vacuum
Schwarzschild blackhole of equal horizon area (equivalently, of
equal entropy). Specifically, the inequality
\begin{equation}
k T_H \leq {\hbar\over4\pi r_H} \equiv {\hbar\over\sqrt{4\pi A_H}}
\end{equation}
appears to be satisfied.

I claim that this inequality is not an accident, but rather that
this inequality is related to the {\sl classical} nature of the
fields interacting with the blackhole. Indeed it shall be shown
that, for a general spherically symmetric distribution of matter
with a blackhole at the center, the Hawking temperature is given
by
\begin{equation}
k T_H = {\hbar\over4\pi r_H} \; e^{-\phi(r_H)} \;
        \left( 1 - 8\pi G \; \rho_H \; r_H^2 \right).
\end{equation}
Now $r_H$ and $\rho_H$, the radius and matter density at the horizon,
clearly depend only on conditions local to the horizon itself. In
contrast, $\phi(r_H)$ is an integral quantity that depends on the
distribution of matter all the way from $r=r_H$ to $r=\infty$. The
remarkable feature of the analysis is that, if the matter surrounding
the blackhole satisfies the Weak Energy Condition (WEC), which is
certainly the case for classical matter, then the Einstein field
equations imply that $\phi(r_H)$ is non--negative. The inequality
$k T_H \leq \hbar/(4\pi r_H)$ follows immediately.

(\underbar{Warning:} Since semiclassical quantum effects are capable
of violating the WEC, it follows that quantum physics may allow a
violation of this inequality. On the other hand, violations of the
WEC in the vicinity of the event horizon are quite likely to
destabilize the horizon, disrupt the blackhole, and lead to a
traversable wormhole, thereby rendering moot the question of the
Hawking temperature \cite{Morris-Thorne}.)

A side effect of the investigation is the discovery of a particularly
pleasant functional parameterization of the static spherically
symmetric metric that permits a simple (formal) integration of the
Einstein field equations in a form suitable for the direct application
of the WEC.

Also of note is the fact that the matter fields at the horizon (as
measured by a fiducial observer --- a FIDO) are constrained to
satisfy the boundary condition $\rho_H = \tau_H$ if the horizon is
to be ``canonical'' in a sense to be described below. This boundary
condition is in fact equivalent to demanding that the energy density
measured by a freely falling observer (FFO) remain integrable as the
observer crosses the horizon.

Several examples are discussed in detail: The Reissner--Nordstrom
geometry and a ``thin shell'' example are particularly instructive
elementary examples. The dyonic dilatonic blackholes and their ilk
are decidedly nontrivial examples.

Finally a conjecture is formulated as to a possible generalization
of these results to spherically asymmetric non-static dirty
blackholes.

\underbar{Units:} Adopt units where $c\equiv 1$, but all other
quantities retain their usual dimensionalities, so that in particular
$G\equiv \ell_P/m_P \equiv \hbar/m_P^2 \equiv \ell_P^2/\hbar$.

%%%%%%%%%%%%%%%%%%%%%%%%%%%%%%%%%%%%%%%%%%%%%%%%%%%%%%%%%%%%%%%%%%%%%%%%%%%%%%
\newpage
%%%%%%%%%%%%%%%%%%%%%%%%%%%%%%%%%%%%%%%%%%%%%%%%%%%%%%%%%%%%%%%%%%%%%%%%%%%%%%
\section{METRIC}
%%%%%%%%%%%%%%%%%%%%%%%%%%%%%%%%%%%%%%%%%%%%%%%%%%%%%%%%%%%%%%%%%%%%%%%%%%%%%%
\subsection{Functional form}
%%%%%%%%%%%%%%%%%%%%%%%%%%%%%%%%%%%%%%%%%%%%%%%%%%%%%%%%%%%%%%%%%%%%%%%%%%%%%%

The spacetime metric generated by any static spherically symmetric
distribution of matter may (without loss of generality) be cast
into the form
\begin{equation}
ds^2 = - g_{tt} \; dt^2 + g_{rr} \; dr^2
       + r^2(d\theta^2 + \sin^2\theta \; d\varphi^2).
\end{equation}
This form corresponds to the adoption of Schwarzschild coordinates.
While one can relatively easily adopt the brute force approach of
inserting this metric into the curvature computation formalism and
``turning the crank'', the resulting expression for the Einstein
tensor is not as illuminating as it might otherwise be.

There is an art to further specifying the functional form of $g_{tt}$
and $g_{rr}$ in such a manner as to keep computations (and their
interpretations) simple. For instance, to discuss traversable
wormholes Morris and Thorne found the choices $g_{tt} = \exp(2\phi(r))$;
$g_{rr} = (1-b(r)/r)^{-1}$ to be particularly advantageous
\cite{Morris-Thorne}. For the discussion currently at hand I propose
\begin{equation}
g_{tt} = e^{-2\phi(r)} \left( 1 - {b(r)\over r} \right), \qquad
g_{rr} = \left( 1 - {b(r)\over r} \right)^{-1}.
\end{equation}
That is:
\begin{equation}
ds^2 = - e^{-2\phi(r)} \left(1 - b(r)/r\right)dt^2
       + {dr^2\over\left( 1 - b(r)/r \right)}
       + r^2(d\theta^2 + \sin^2\theta \; d\varphi^2).
\label{metric}
\end{equation}
Following Morris and Thorne, the function $b(r)$ will be referred
to as the ``shape function''. The shape function may be thought of
as specifying the shape of the spatial slices.  On the other hand,
$\phi(r)$ might best be interpreted as a sort of ``anomalous
redshift'' that describes how far the total gravitational redshift
deviates from that implied by the shape function. As will subsequently
be seen the Einstein field equations have a particularly nice form
when written in terms of these functions.

%%%%%%%%%%%%%%%%%%%%%%%%%%%%%%%%%%%%%%%%%%%%%%%%%%%%%%%%%%%%%%%%%%%%%%%%%%%%%%
\subsection{Putative horizons}
%%%%%%%%%%%%%%%%%%%%%%%%%%%%%%%%%%%%%%%%%%%%%%%%%%%%%%%%%%%%%%%%%%%%%%%%%%%%%%

For now, explore the meaning of the metric in the form
(\ref{metric}) without yet applying the field equations. Firstly,
applying boundary conditions at spatial infinity permits one to
set $\phi(\infty) = 0$ without loss of generality. Once this
normalization of the asymptotic time coordinate is adopted one may
interpret $b(\infty)$ in terms of the asymptotic mass $b(\infty)
= 2GM$.  (Naturally one is assuming an asymptotically flat geometry).

The metric (\ref{metric}) has putative horizons at values of $r$
satisfying $b(r_H) = r_H$. Only the outermost horizon is of immediate
interest and comments will be restricted to that case. Now for the
outermost horizon one has $\forall r > r_H$ that $b(r) < r$,
consequently $b'(r_H) \leq 1$. The case $b'(r_H) = 1$ is anomalous
and will be discussed separately. Assuming then that $b'(r_H) < 1$
the behaviour of the metric near the putative horizon is
\begin{equation}
ds^2 \approx - e^{-2\phi(r_H)} \left({r-r_H\over r_H}\right)
                               (1-b'(r_H)) dt^2
       + {1\over(1-b'(r_H))} \left({r_H\over r-r_H}\right) dr^2
       + r_H^2(d\theta^2 + \sin^2\theta \; d\varphi^2).
\end{equation}
Thus the putative horizon is seen to possess all the usual properties
of a Schwarzschild horizon provided that $e^{-2\phi(r)}$ is positive
and of finite slope at $r=r_H$, corresponding to $|\phi(r_H)|$ and
$|\phi'(r_H)|$ being finite.

The putative horizon at $r_H = b(r_H)$ will be said to be of {\it
canonical type} if
\begin{equation}
b'(r_H) <1; \qquad |\phi(r_H)| < \infty; \qquad |\phi'(r_H)| < \infty.
\end{equation}

Noncanonical horizons are of interest in their own right. On the
one hand, if $b'(r_H) = 1$ one may Taylor expand
\begin{eqnarray}
b(r) =&& b(r_H) + b'(r_H) (r-r_H) + {b''(r_H)\over2} (r-r_H)^2 + ...
         \nonumber\\
     =&& r + {\gamma_2\over2 r_H} (r-r_H)^2 + ...
\end{eqnarray}
This allows the simple expansion $(1-{b/r}) = {1\over2}\gamma_2
(r-r_H)^2/r_H^2 + ...$, thus indicating that in this case $g_{rr}$
does not change sign at the horizon (provided that $\gamma_2 \neq
0$).  This behaviour is an indication of the merging of an inner
and an outer horizon. In fact, the horizon of an extreme $Q=M$
Reissner--Nordstrom blackhole is precisely of this type (with
$\phi(r)\equiv 0$). If $\gamma_2 = 0$ then one must go to higher
order in the Taylor series expansion. If the first nonzero term is
of oder $n$, that is if $b(r) - r = {1\over n!}\gamma_n (r-r_H)^n/r_H^n
+ ... $, then one may easily convince oneself that one is dealing
with a $n$--fold merging of $n$ degenerate horizons.

On the other hand, even if $b'(r_H) <1$, one may still obtain
noncanonical horizon structure due to the behaviour of $\phi(r)$
near the putative horizon. For instance, take $\phi(r) = +{1\over2}
\ln({r-r_H\over r_H}) + f(r)$, where $f(r)$ is smooth and finite
at the putative horizon. In this case the behaviour of the metric
near the putative horizon is
\begin{equation}
ds^2 \approx - e^{-2 f(r_H)} (1-b'(r_H)) dt^2
       + {1\over(1-b'(r_H))} \left({r_H\over r-r_H}\right) dr^2
       + r_H^2(d\theta^2 + \sin^2\theta \; d\varphi^2).
\end{equation}
Thus $g_{tt}$ remains nonzero on the putative horizon, so that the
putative horizon is not in fact a horizon at all, but rather is
the throat of a traversable wormhole \cite{Morris-Thorne}.

Finally, one should consider the possibility that the ``anomalous
redshift'' might diverge in a region where the ``shape function''
is still well behaved. Specifically, consider the possibility that
$\phi(r) \to +\infty$ as $r\to r_H$, while $b(r)\to r_0 \equiv 2
G m_0 < r_H$. Such a horizon is certainly noncanonical. Analysis
of the Einstein field equations (see below) indicates that this
case corresponds to a divergence in the stress--energy density as
the horizon is approached.

Further discussion of noncanonical horizons will be postponed, and
henceforth all horizons are taken to be of canonical type.

%%%%%%%%%%%%%%%%%%%%%%%%%%%%%%%%%%%%%%%%%%%%%%%%%%%%%%%%%%%%%%%%%%%%%%%%%%%%%%
\newpage
%%%%%%%%%%%%%%%%%%%%%%%%%%%%%%%%%%%%%%%%%%%%%%%%%%%%%%%%%%%%%%%%%%%%%%%%%%%%
\section{HAWKING TEMPERATURE}
%%%%%%%%%%%%%%%%%%%%%%%%%%%%%%%%%%%%%%%%%%%%%%%%%%%%%%%%%%%%%%%%%%%%%%%%%%%%%%
\subsection{Surface gravity}
%%%%%%%%%%%%%%%%%%%%%%%%%%%%%%%%%%%%%%%%%%%%%%%%%%%%%%%%%%%%%%%%%%%%%%%%%%%%%%

The Hawking temperature of a blackhole is given in terms of its
surface gravity by $k T_H = (\hbar/2\pi) \kappa$.  Now in general
for a spherically symmetric system the surface gravity can be
computed via
\begin{equation}
\kappa = \lim_{r\to r_H} \left\{ {1\over2}
                         {\partial_r g_{tt} \over\sqrt{g_{tt} g_{rr}}}
	                 \right\}_.
\end{equation}
(This result holds independently of whether or not one chooses to
normalize the $g_{\theta\theta}$ and $g_{\varphi\varphi}$ components
of the metric by adopting Schwarzschild coordinates.) For the choice
of functional form described in (\ref{metric}) this implies
\begin{eqnarray}
\kappa =&& \lim_{r\to r_H}
           \left\{ {1\over2} \; e^\phi \;
                   {\partial\over\partial r} \left[ e^{-2\phi}
	           \left(1-{b(r)\over r} \right) \right]
           \right\} \nonumber \\
       =&& \lim_{r\to r_H}
           \left\{ {1\over2} \; e^{-\phi} \;
	            \left[ -2\phi'(r) \left( 1 - {b(r)\over r} \right)
		           + {b(r)\over r^2} - {b'(r)\over r}
		    \right].
	   \right\}
\end{eqnarray}
Now for a canonical horizon $|\phi(r_H)|$ and $|\phi'(r_H)|$ are
both finite so that
\begin{equation}
\kappa = {1\over2 r_H} e^{-\phi(r_H)} \left( 1-b'(r_H) \right).
\end{equation}
At this stage of course, this formula is largely {\it definition}.
This formula receives its physical significance only after $b'(r_H)$
and $\phi(r_H)$ are related to the distribution of matter by imposing
the Einstein field equations. Note that the derivation of the
formula for the surface gravity continues to make perfectly good
sense for degenerate horizons ({\it ie} $b'(r_H)=1$), merely
asserting in this case that $\kappa = 0$.

%%%%%%%%%%%%%%%%%%%%%%%%%%%%%%%%%%%%%%%%%%%%%%%%%%%%%%%%%%%%%%%%%%%%%%%%%%%%%%
\subsection{Euclidean signature techniques}
%%%%%%%%%%%%%%%%%%%%%%%%%%%%%%%%%%%%%%%%%%%%%%%%%%%%%%%%%%%%%%%%%%%%%%%%%%%%%%

Another way of calculating the Hawking temperature is via the
periodicity of the Euclidean signature analytic continuation of
the manifold \cite{incantations}. Proceed by making the formal
substitution $t\to -it$ to yield
\begin{equation}
ds^2_E = + e^{-2\phi(r)} \left(1 - b(r)/r\right) dt^2
       + {dr^2\over\left( 1 - {b(r)/r} \right)}
       + r^2(d\theta^2 + \sin^2\theta d\varphi^2).
\end{equation}
As is usual, discard the entire $r<r_H$ region, retaining only the
(analytic continuation of) that region that was outside the outermost
horizon ({\it ie:} $r\geq r_H$). Taylor series expand the metric
in the region $r\approx r_H$. Provided that the horizon is canonical
one may write $(1-{b/r}) \equiv (r-b)/r \approx (r-r_H) r_H^{-1}
(1-b'(r_H))$ to give
\begin{equation}
ds^2_E \approx - e^{-2\phi(r_H)} (1-b'(r_H))
                 \left({r-r_H\over r_H}\right)   dt^2
       + {1\over(1-b'(r_H))} \left({r_H\over r-r_H}\right) dr^2
       + r_H^2(d\theta^2 + \sin^2\theta\; d\varphi^2).
\end{equation}
Construct a new radial variable $\varrho$ by taking
\begin{equation}
d\varrho = {1\over\sqrt{1-b'(r_H)}}\sqrt{{r_H\over r-r_H}} dr
         = {2\over\sqrt{1-b'(r_H)}}d(\sqrt{r_H(r-r_H)}).
\end{equation}
Then $r_H(r-r_H) = {1\over4} (1-b'(r_H)) \varrho^2$, and the
Euclidean signature metric may be written as
\begin{equation}
ds^2_E \approx - e^{-2\phi(r_H)} (1-b'(r_H))^2 {1\over 4 r_H^2} \;
 (\varrho^2 \; dt^2)
       + d\varrho^2
       + r_H^2(d\theta^2 + \sin^2\theta \; d\varphi^2).
\end{equation}
Now the $(\varrho,t)$ plane is a smooth two dimensional manifold
if and only if $t$ is interpreted as an angular variable with period
\begin{equation}
\beta = 2\pi \;2 r_H \; e^\phi(r_H) \;(1-b'(r_H))^{-1}.
\end{equation}
Invoking the usual incantations \cite{incantations}, this periodicity
in imaginary (Euclidean) time is interpreted as evidence of a
thermal bath of temperature $k T = \hbar /\beta$, so that the
Hawking temperature is identified as
\begin{equation}
k T_H = {\hbar\over 4\pi r_H} \; e^{-\phi(r_H)} \; (1-b'(r_H)).
\end{equation}
This is the same result as was obtained by direct calculation of
the surface gravity, though this formulation has the advantage of
(1) shedding further illumination on the subtleties associated with
noncanonical horizons, and (2) verifying the relationship between
Hawking temperature and surface gravity.

%%%%%%%%%%%%%%%%%%%%%%%%%%%%%%%%%%%%%%%%%%%%%%%%%%%%%%%%%%%%%%%%%%%%%%%%%%%%%%
\newpage
%%%%%%%%%%%%%%%%%%%%%%%%%%%%%%%%%%%%%%%%%%%%%%%%%%%%%%%%%%%%%%%%%%%%%%%%%%%%%%
\section{EINSTEIN FIELD EQUATIONS}
%%%%%%%%%%%%%%%%%%%%%%%%%%%%%%%%%%%%%%%%%%%%%%%%%%%%%%%%%%%%%%%%%%%%%%%%%%%%%%
\subsection{Formal solution}
%%%%%%%%%%%%%%%%%%%%%%%%%%%%%%%%%%%%%%%%%%%%%%%%%%%%%%%%%%%%%%%%%%%%%%%%%%%%%%

The Einstein tensor corresponding to (\ref{metric}) can be obtained
by the standard simple but tedious computation. Choose an orthonormal
basis attached to the $(t,r,\theta,\varphi)$ coordinate system
({\it ie}, choose a fiducial observer basis --- a FIDO basis)
\begin{eqnarray}
G_{\hat t \hat t} =&& {b'\over r^2} \\
G_{\hat r \hat r} =&& -{2\over r} \left( 1-{b\over r} \right) \phi'
                      - {b'\over r^2}
\end{eqnarray}
Whereas the forms of $G_{\hat t \hat t}$ and $G_{\hat r \hat r}$
are quite pleasing, the form of $G_{\hat\theta\hat\theta} \equiv
G_{\hat\varphi\hat\varphi}$ is quite horrible. Fortunately one will
not need to use $G_{\hat\theta\hat\theta}$ or $G_{\hat\varphi\hat\varphi}$
explicitly. For completeness note:
\begin{eqnarray}
G_{\hat\theta\hat\theta} = G_{\hat\varphi\hat\varphi}
  =&& \left( 1- {b\over r}\right)
      \left(-\phi'' +\phi' \left(\phi'-{1\over r}\right) \right)
      \nonumber \\
   &&-{3\over2}\phi'\left({b\over r^2}-{b'\over r}\right)
      -{1\over2}{b''\over r}.
\end{eqnarray}
All other components of the Einstein tensor are zero.  To minimize
computation use the results of Morris and Thorne \cite{Morris-Thorne}
with the substitution $\phi_{\rm Morris-Thorne} = -\phi_{\rm here}
+ {1\over2}\ln(1-{b/r})$.

The Einstein field equations are
\begin{equation}
G_{\alpha\beta} = 8\pi G \; T_{\alpha\beta}
                = 8\pi \; {\ell_P^2\over\hbar} \; T_{\alpha\beta}.
\end{equation}
In the FIDO orthonormal basis used above, the nonzero components of the
stress--energy tensor are
\begin{equation}
T_{\hat t \hat t} = \rho; \qquad T_{\hat r \hat r} = -\tau; \qquad
T_{\hat\theta\hat\theta} = T_{\hat\varphi\hat\varphi} = p.
\end{equation}
The first two Einstein equations are then simply rearranged to give
\begin{eqnarray}
b' =&& 8\pi G \; \rho \; r^2, \\
\phi' =&& - {8\pi G\over2} {(\rho-\tau)r\over(1-b/r)}.
\label{einstein}
\end{eqnarray}
Instead of imposing the third Einstein equation $G_{\hat\theta\hat\theta}
= G_{\hat\varphi\hat\varphi} = 8 \pi\; G\; p$, observe that (as is
usual) this equation is redundant with the imposition of the
conservation of stress--energy. Thus one may take the third equation
to be
\begin{equation}
\tau' = (\rho-\tau)[-\phi' +{1\over2}\{\ln(1-b/r)\}'] - 2(p+\tau)/r.
\end{equation}
Taking $\rho$ and $\tau$ to be primary, one may formally integrate
the Einstein equations, and then substitute this into the conservation
of stress--energy to determine $p$. Specifically:
\begin{eqnarray}
b(r) =&& r_H + 8\pi G \int_{r_H}^r \rho {\tilde r}^2 d{\tilde r}, \\
\phi(r) =&& {8\pi G\over 2}
            \int_r^\infty {(\rho-\tau)\tilde r\over(1-b/{\tilde r})}
	    d{\tilde r}, \\
p(r) =&& {r\over2}
      \left[{(\rho-\tau)\over2(1-b/r)}
            \left\{{ b-8\pi G \;\tau \; r^3 \over r^2} \right\} -\tau'
      \right] -\tau.
\end{eqnarray}
Inserting these results into the formula for the Hawking temperature
now yields the promised result
\begin{equation}
k T_H = {\hbar\over4\pi r_H} \;
        \exp\left(-{8\pi G\over 2}
	         \int_{r_H}^{\infty}{(\rho-\tau)r\over(1-b/r)} dr
	    \right) \;
        \left( 1 - 8\pi G \;\rho_H \; r_H^2 \right).
\end{equation}
The Hawking temperature is seen to depend both on data local to
the event horizon $(r_H,\rho_H)$ and on a ``redshift'' factor whose
computation requires knowledge of $\rho(r)$ and $\tau(r)$ all the
way from the horizon to spatial infinity.

Once the problem has been cast in this form the role of the Weak
Energy Condition is manifest. WEC implies that $\rho - \tau \geq 0$
and that $\rho\geq0$. Consequently $\forall r$, $\phi(r)\geq 0$.
Also $b'(r_H) \geq 0$. Thus adopting WEC allows one to assert the
promised inequality
\begin{equation}
k T_H \leq {\hbar\over4\pi r_H} \equiv {\hbar\over\sqrt{4\pi A_H}}.
\end{equation}

%%%%%%%%%%%%%%%%%%%%%%%%%%%%%%%%%%%%%%%%%%%%%%%%%%%%%%%%%%%%%%%%%%%%%%%%%%%%%%
\subsection{Convergence issues}
%%%%%%%%%%%%%%%%%%%%%%%%%%%%%%%%%%%%%%%%%%%%%%%%%%%%%%%%%%%%%%%%%%%%%%%%%%%%%%

Several points regarding these formulae are worth mentioning.
Firstly, the condition $b'(r_H)\leq1$ which is automatically
satisfied by the outermost putative horizon (regardless of whether
or not it be canonical) implies, via the Einstein field equations,
a constraint on $\rho_H$, {\it viz} $\rho_H < 1/(8\pi G r_H^2)
\equiv \hbar/(8\pi\ell_P^2 r_H^2)$. This constraint has the nice
feature of guaranteeing that the Hawking temperature is non-negative.
Turning to questions of convergence of the various integrals
encountered, note that
\begin{equation}
2GM = r_H + 2G \int_{r_H}^\infty 4\pi\rho r^2 dr,
\end{equation}
so that this integral is guaranteed to converge by the assumed
asymptotic flatness of the spacetime. The only questionable integral
is that for $\phi(r_H)$. Specifically, its convergence properties
near the putative horizon are somewhat subtle. Assuming $b'(r_H)<1$
one may write this integral as
\begin{eqnarray}
\phi(r_H) &&\equiv  {8\pi G\over 2}
            \int_{r_H}^\infty {(\rho-\tau)r\over(1-b/r)} dr,
	    \nonumber \\
	  &&\approx ({\rm finite}) + {8\pi G r_H^2\over 2(1-b'(r_H))}
            \int_{r_H}^{(1+\epsilon)r_H} {(\rho-\tau)\over(r-r_H)} dr.
\end{eqnarray}
This integral converges provided that $(\rho-\tau) \leq k(r-r_H)^\alpha$
as $r\to r_H$ for some arbitrary constant $k$ and some constant
$\alpha>0$. In particular this implies that $\rho_H=\tau_H$ is a
necessary condition for the existence of a canonical horizon. It
should come as no great surprise then to observe that all ``reasonable''
classical field solutions satisfy this boundary condition. Indeed,
this boundary condition is equivalent to requiring the energy
density measured by a freely falling observer (FFO) to remain
integrable as one crosses the horizon.

To see this, consider a freely falling observer who starts falling
from spatial infinity with initial velocity zero. Let $V^\mu$ denote
the four--velocity of the FFO, and let $K^\mu$ denote the timelike
Killing vector. That is, $K^\mu = (1,0,0,0)$; $K_\mu=(-g_{tt},0,0,0)$.
Then the inner product $K^\mu V_\mu$ is conserved along geodesics,
so that $V_t = 1$, $V^t = -g^{tt} = -1/g_{tt}$. Since the four--velocity
must be normalized ($\Vert V \Vert = -1$), one may solve for the
radial component to find (outside the outermost horizon):
\begin{equation}
V^\mu = ({1\over g_{tt}},
         \sqrt{ 1 \over g_{rr} } \sqrt{ {1 \over g_{tt}} - 1},
	 0,0).
\end{equation}
In the FIDO basis
\begin{equation}
V^{\hat\mu} = ( \sqrt{ 1 \over g_{tt} },
                \sqrt{ {1 \over g_{tt}} - 1}, 0,0).
\end{equation}
So the energy density measured by a FFO is $\rho_{\rm FFO} \equiv
T_{\hat\mu \hat\nu} V^{\hat\mu} V^{\hat\nu} = \rho/g_{tt} +
(-\tau)(g_{tt}^{-1}-1) = \tau + (\rho-\tau)/g_{tt}$. Finally,
inserting the functional form for $g_{tt}$ one sees
\begin{equation}
\rho_{\rm FFO} = \tau + {(\rho-\tau)\over e^{-2\phi} (1-b/r)} \approx
             {e^{+2\phi} (\rho-\tau) r_H \over (1-b')(r-r_H)}.
\end{equation}
So that the boundary condition $(\rho-\tau)\leq k(r-r_H)^\alpha$,
$\alpha>0$, required to keep $\phi(r_H)$ finite, implies the
integrability of $\rho_{\rm FFO}$. Conversely, the integrability of
$\rho_{FFO}$ implies either (1) the finiteness of $\phi(r_H)$
(canonical horizon), or (2) $\phi(r)\to -\infty$ (corresponding to
a traversable wormhole).

%%%%%%%%%%%%%%%%%%%%%%%%%%%%%%%%%%%%%%%%%%%%%%%%%%%%%%%%%%%%%%%%%%%%%%%%%%%%%%
\newpage
%%%%%%%%%%%%%%%%%%%%%%%%%%%%%%%%%%%%%%%%%%%%%%%%%%%%%%%%%%%%%%%%%%%%%%%%%%%%%%
\section{EXAMPLES}
%%%%%%%%%%%%%%%%%%%%%%%%%%%%%%%%%%%%%%%%%%%%%%%%%%%%%%%%%%%%%%%%%%%%%%%%%%%%%%
\subsection{Reissner--Nordstrom}
%%%%%%%%%%%%%%%%%%%%%%%%%%%%%%%%%%%%%%%%%%%%%%%%%%%%%%%%%%%%%%%%%%%%%%%%%%%%%%

For the Reissner--Nordstrom geometry the symmetries of the situation
together with the form of the electromagnetic stress--energy tensor
implies
\begin{equation}
\rho = \tau = p = E^2/8\pi.
\end{equation}
This automatically gives $\phi(r)=0$, $\forall r$. The electromagnetic
field equations imply $E = Q/r^2$, so that
\begin{equation}
k T_H^{\rm RN} = {\hbar\over4\pi r_H}
           \left(1 - {GQ^2\over r_H^2} \right).
\end{equation}
This is an unusual, though correct formula for the Hawking temperature
of a Reissner--Nordstrom blackhole. To see this note that explicit
solution of the Einstein--Maxwell field equations gives $g_{tt} =
(g_{rr})^{-1} = 1 - (2GM/r) + (GQ^2/r^2)$, whence $\kappa = {1\over2}
\lim_{r\to r_H} \partial_r g_{tt} = {1\over2} (\{2GM/r_H^2\} -
\{2GQ^2/r_H^2\}) = (1/2r_H)( \{2GM/r_H\} - \{GQ^2/r_H^2\}) = (1/2r_H)
(1-\{GQ^2/r_H^2\})$, which is the above result.

%%%%%%%%%%%%%%%%%%%%%%%%%%%%%%%%%%%%%%%%%%%%%%%%%%%%%%%%%%%%%%%%%%%%%%%%%%%%%%
\subsection{Thin shell geometry}
%%%%%%%%%%%%%%%%%%%%%%%%%%%%%%%%%%%%%%%%%%%%%%%%%%%%%%%%%%%%%%%%%%%%%%%%%%%%%%

Consider a thin spherical shell of matter of density $\rho_S$,
radius $r_S$, and thickness $(\delta r)_S$, which surrounds a vacuum
blackhole of Schwarzschild radius $r_H$. The mass of this thin
shell is $m_S = 4\pi \rho_S r_S^2 (\delta r)_S$, and the asymptotic
total mass satisfies $2 G M = r_H + 2 G m_S$. The shape function exhibits
a step function discontinuity: $b(r) = r_H + \Theta(r-r_S) 2 G m_S$.
Direct integration of $\phi'(r)$ is not an appropriate way of
calculating $\phi(r_H)$ due to the discontinuity in $b(r)$. Rather
it is more appropriate to solve for $\phi(r_H)$ by using the
continuity of $g_{tt}$ to develop matching conditions.  Everywhere
except at the shell itself both $\rho$ and $\tau$ are zero, so
$\phi(r)$ is piecewise constant.  Applying boundary conditions at
the horizon and at spatial infinity gives $\phi(r) = \phi(r_H)
\Theta(r_S-r)$. The matching conditions are thus
\begin{eqnarray}
g_{tt}(r_S^+) =&& 1- 2GM/r_S, \\
g_{tt}(r_S^-) =&& e^{-2\phi(r_H)} \; (1- r_H/r_S).
\end{eqnarray}
One immediately obtains
\begin{equation}
e^{-2\phi(r_H)} = { {1- 2GM/r_S} \over {1- r_H/r_S} }
               = 1 - { {2Gm_S/r_S} \over {1-r_H/r_S}}.
\end{equation}
Finally, noting that $\rho=0$ on the horizon, one sees that the
Hawking temperature is suppressed by
\begin{equation}
k T_H = {\hbar\over4\pi r_H}
           \sqrt{1 - {2Gm_S/r_S\over(1-r_H/r_S)} }.
\end{equation}
Physically, this suppression of the Hawking temperature may be
attributed to the fact that the shell introduces an extra gravitational
redshift that decreases the energy of the Hawking photons on their
way out to spatial infinity.

%%%%%%%%%%%%%%%%%%%%%%%%%%%%%%%%%%%%%%%%%%%%%%%%%%%%%%%%%%%%%%%%%%%%%%%%%%%%%%
\subsection{Charged dilatonic blackhole}
%%%%%%%%%%%%%%%%%%%%%%%%%%%%%%%%%%%%%%%%%%%%%%%%%%%%%%%%%%%%%%%%%%%%%%%%%%%%%%

As a decidedly nontrivial example consider geometry and fields
surrounding a charged dilatonic blackhole \cite{Gibbons-Maeda,GHS}.
The calculation about to be exhibited is a rather obtuse way of
calculating the Hawking temperature, depending as it does on delicate
cancellations amoung $r_H$, $\rho_H$, and $\phi_H$.  The only virtue
of this computation is that it illustrates general features of the
formalism. (\underbar{Units:} For this section only set $G\equiv 1$.)

Consider then a solution to the combined (gravity + electromagnetism
+ dilaton) equations of motion. The Lagrangian is
\begin{equation}
{\cal L}=\sqrt{-g}\left\{-R/8\pi + 2 (\nabla\Phi)^2 + F^2/4\pi \right\}.
\end{equation}
(\underbar{Warning:} $\Phi \neq \phi$!)
In Schwarzschild coordinates the solution corresponding to an
electric monopole is
\begin{eqnarray}
ds^2 = &&- \left( 1 - {2M\over a +\sqrt{r^2+a^2} } \right) dt^2
         + \left( 1 - {2M\over a +\sqrt{r^2+a^2} } \right)^{-1}
          {r^2\over r^2+a^2}dr^2
	  \nonumber \\
       &&+ r^2(d\theta^2 + \sin^2\theta \; d\varphi^2), \\
F_{\hat t \hat r} = &&Q/r^2, \\
e^{2\Phi} = &&1 - {Q^2\over M(a+\sqrt{r^2+a^2})}
\end{eqnarray}
Here one has used the freedom to make an overall shift in $\Phi$
to set $\Phi(\infty)=0$. The parameter $a$ is defined by $a\equiv
Q^2/2M$. In terms of the formalism developed in this paper
\begin{eqnarray}
1-{b\over r} &&= \left( 1 + {a^2\over r^2} \right)
                 \left( 1 - {2M\over a +\sqrt{r^2+a^2} } \right), \\
e^{-2\phi(r)} &&= \left( 1 + {a^2\over r^2} \right)^{-1}
                = {r^2\over r^2 + a^2}.
\end{eqnarray}
The horizon occurs at $2M=a+\sqrt{r_H^2+a^2}$, that is, $r_H^2 +
a^2 = (2M-a)^2$, so that the surface gravity is
\begin{equation}
\kappa = {1\over2 r_H} \; {r_H\over\sqrt{r_H^2 + a^2}} \;
           \left( 1 - 8\pi \;\rho_H \; r_H^2 \right)
       = {1\over2(2M-a)}  \left( 1 - 8\pi \;\rho_H \; r_H^2 \right).
\end{equation}
To calculate $\rho_H$ one evaluates the nonzero components of the
stress-energy tensor
\begin{eqnarray}
\rho =&& {1\over8\pi} e^{-2\Phi} E^2 + \Vert\nabla\Phi\Vert^2, \\
\tau =&& {1\over8\pi} e^{-2\Phi} E^2 - \Vert\nabla\Phi\Vert^2, \\
p    =&& {1\over8\pi} e^{-2\Phi} E^2 - \Vert\nabla\Phi\Vert^2.
\end{eqnarray}
As one approaches the event horizon it is easy to verify that $\Vert
\nabla\Phi \Vert \to 0$, while $E \to Q/r_H^2$, so that $\rho \to
{1\over8\pi} (1 - \{Q^2/2M^2\} ) (Q^2/r_H^4)$. Thus
\begin{equation}
 8\pi \;\rho_H \; r_H^2
 = \left(1-{Q^2\over2M^2}\right){Q^2\over r_H^2}
 = {M-a\over M} {Q^2\over2M(2M-2a)}
 = {a\over2M}.
\end{equation}
Combining this considerable morass yields the simple result
\begin{equation}
\kappa = {1\over4M}
\end{equation}
As previously mentioned, this calculation is a particularly obtuse
manner in which to compute the surface gravity. This computation
is of interest only insofar as it illustrates general principles
and serves as a check on the formalism.  The inequality $\kappa <
1/(2r_H)$, which previously appeared to be just a random accident
of the calculation, is now seen to be intrinsically related to the
fact that classical fields satisfy the WEC.

%%%%%%%%%%%%%%%%%%%%%%%%%%%%%%%%%%%%%%%%%%%%%%%%%%%%%%%%%%%%%%%%%%%%%%%%%%%%%%
\newpage
%%%%%%%%%%%%%%%%%%%%%%%%%%%%%%%%%%%%%%%%%%%%%%%%%%%%%%%%%%%%%%%%%%%%%%%%%%%%%%
\section{DISCUSSION}
%%%%%%%%%%%%%%%%%%%%%%%%%%%%%%%%%%%%%%%%%%%%%%%%%%%%%%%%%%%%%%%%%%%%%%%%%%%%%%

For an arbitrary static spherically symmetric blackhole this note
has established a general formula for the Hawking temperature in
terms of the energy density and radial tension. Adopting Schwarzschild
coordinates, and writing
\begin{equation}
b(r) = r_H + 8\pi G \int_{r_H}^r \rho {\tilde r}^2 d{\tilde r},
\end{equation}
one finds that
\begin{equation}
k T_H = {\hbar\over4\pi r_H} \;
        \exp\left(-{8\pi G\over 2}
	         \int_{r_H}^{\infty}{(\rho-\tau)r\over(1-b/r)} dr
	    \right) \;
        \left( 1 - 8\pi G \;\rho_H \; r_H^2 \right).
\end{equation}
Generalizations of this result to axisymmetric spacetimes (for
instance, to Kerr--Newman blackholes embedded in an axisymmetric
cloud of matter) would clearly be of interest. Generalizations to
arbitrary event horizons are probably unmanageable. On the one
hand, the Dominant Energy Condition (DEC) guarantees the constancy of
the surface gravity (and hence the constancy of the Hawking
temperature) over the surface of an arbitrary stationary event horizon.
Furthermore, one might conceivably hope to generalize the factor
$4\pi r_H$ to $\sqrt{4\pi A_H}$. On the other hand, there is no
particular reason to believe that $\rho_H$ is constant over the
event horizon, nor is it clear how to generalize the notion of
$\phi(r_H)$. (Presumably in terms of some line integral from the
horizon to spatial infinity?)

If the central result of this paper is supplemented by the Weak
Energy Condition one may further assert (for static spherically
symmetric dirty blackholes) the general inequality
\begin{equation}
k T_H \leq {\hbar\over4\pi r_H}.
\end{equation}
This inequality may be somewhat strengthened if one explicitly
separates out the electromagnetic contribution to the stress--energy.
Note that $\rho_H \geq (\rho_{\rm em})_H \equiv E^2/8\pi \equiv
Q^2/(8\pi r_H^4)$. Thus for electrically charged static spherically
symmetric dirty blackholes
\begin{equation}
k T_H \leq {\hbar\over4\pi r_H}
           \left(1 - {GQ^2\over r_H^2} \right).
\end{equation}
(Generalization to magnetic charge and the dyonic case is trivial.)
The possibility of further generalizing these inequalities is more
promising. I will restrain myself to a single\\
\underbar{Conjecture:}\\
For a stationary dirty blackhole in interaction with matter fields
satisfying the Dominant Energy Condition
\begin{equation}
k T_H \leq {\hbar\over\sqrt{4\pi A_H}}.
\end{equation}
\underbar{Notes:} (1) It should be noted that this inequality
is satisfied by the Kerr--Newman geometry. (2) The restrictions
``stationary'' and ``Dominant Energy Condition'' cannot be dispensed
with as they are required merely in order to guarantee the existence
of a constant Hawking temperature.  (3) With regard to this
conjectured inequality, it should be pointed out that a weaker
inequality that requires stronger hypotheses can be derived from
the ``four laws of blackhole mechanics'' \cite{Bardeen-Carter-Hawking}.
Restricting the results of that paper to the case of zero rotation,
one observes the equality ($S_H = {\rm entropy} = (1/4) k A_H
/\ell_P^2$):
\begin{equation}
M = \int_{r_H}^\infty (2T^\mu{}_\nu - T \delta^\mu{}_\nu) K^\nu d\Sigma_\mu
    + 2 T_H S_H.
\end{equation}
By invoking the Strong Energy Condition, the integral can be made
positive, in which case one obtains the inequality
\begin{equation}
k T_H \leq {M\over 2 S_H /k} \equiv {2M \ell_P^2 \over A_H}.
\end{equation}
When restricted to spherical symmetry this reduces to
\begin{equation}
k T_H \leq {2GM\over r_H} {\hbar \over 4\pi r_H}.
\end{equation}
Which is clearly weaker than the inequalities considered above.

In summary, this paper has exhibited a general formalism for
calculating the surface gravity and Hawking temperature of spherically
symmetric static dirty blackholes. The formalism serves to tie
together a number of otherwise seemingly accidental results scattered
throughout the literature. Clear directions for future research
are indicated.

%%%%%%%%%%%%%%%%%%%%%%%%%%%%%%%%%%%%%%%%%%%%%%%%%%%%%%%%%%%%%%%%%%%%%%%%%%%%%%
\acknowledgements
%%%%%%%%%%%%%%%%%%%%%%%%%%%%%%%%%%%%%%%%%%%%%%%%%%%%%%%%%%%%%%%%%%%%%%%%%%%%%%

This research was supported by the U.S. Department of Energy.

%%%%%%%%%%%%%%%%%%%%%%%%%%%%%%%%%%%%%%%%%%%%%%%%%%%%%%%%%%%%%%%%%%%%%%%%%%%%%%
\newpage
%%%%%%%%%%%%%%%%%%%%%%%%%%%%%%%%%%%%%%%%%%%%%%%%%%%%%%%%%%%%%%%%%%%%%%%%%%%%%%

%%%%%%%%%%%%%%%%%%%%%%%%%%%%%%%%%%%%%%%%%%%%%%%%%%%%%%%%%%%%%%%%%%%%%%%%%%%%%%
\end{document}